# HALOC-AxA: An Area/-Energy-Efficient Approximate Adder for Image Processing Application

Hasnain A. Ziad
*Dept. of Electrical and Computer Engineering*
*Florida Polytechnic University*
Lakeland, FL, USA
hziad7099@floridapoly.edu

Ashiq A. Sakib
*Dept. of Electrical and Computer Engineering*
*Southern Illinois University Edwardsville*
Edwardsville, IL, USA
asakib@siue.edu

*Abstract*— The design of approximate adders has been widely researched to advance energy-efficient hardware for computation intensive multimedia applications, such as image/audio/video processing. Several static and dynamic approximate adders exist in the literature, each of which endeavors to balance the conflicting demands of high performance, computational accuracy, and energy efficiency. This work introduces a novel approximate adder that is more energy- and area-efficient than existing adders, while achieving improved or comparable accuracy, as demonstrated by simulation results. The proposed adder's ability to digitally reconstruct high-quality images is further demonstrated by the deployment of the design for an image processing task.

## I. INTRODUCTION

The energy consumption of modern electronics has experienced a substantial increase as a result of the continuous and aggressive scaling of semiconductor technologies. In the era of big data, this trend is more pronounced in computing systems that process vast quantities of data [1]. With the widespread utilization of battery-operated mobile and portable devices, energy-efficiency has become a critical design constraint, particularly for computationally intensive applications such as multimedia processing, deep learning, data mining, and recognition [2]. For applications that exhibit inherent tolerance to minor computational inaccuracies (e.g., image/video/audio processing), approximate computing has emerged as a promising paradigm to address the conflicting demands of high performance and energy efficiency. The inability of human vision or hearing to detect minor distortions allows for perceptual quality to be maintained in such applications, despite slightly inaccurate computation. By selectively relaxing the accuracy where permissible, approximate computation can achieve substantial reductions in energy utilization, area, and delay. General-purpose processors offer limited benefits from approximation as they prioritize high accuracy. However, ASICs tailored to error-resilient applications — such as image and video processing — are well-positioned to fully exploit this paradigm [3, 4].

The design of efficient approximate adders has been a primary focus for researchers aiming to advance energy-efficient hardware, particularly for power-constrained edge and embedded systems. Existing approximate adders broadly fall into two categories: static approximate adders (SAAs) and dynamic approximate adders (DAAs). While SAAs apply fixed approximation patterns, DAAs provide dynamic approximation based on different inputs, offering potential flexibility but often at the cost of increased design complexity. SAAs are favored when inputs are well-characterized and predictable, enabling reliable energy savings without the overhead associated with DAAs. In this paper, we present a novel SAA design that is both energy and area efficient than existing SAAs. Additionally, the proposed adder is deployed in an image processing application to demonstrate the designs ability to reconstruct high-quality images.

The rest of the paper is organized as follows. Section II briefly discusses some of the relevant approximate adders. Section III details the design and operation of the proposed adder. The simulation results and comparison are presented in Section IV, followed by Section V that concludes the paper.

## II. BACKGROUND: APPROXIMATE ADDERS

Current processing units implement parallel adders, e.g., carry look-ahead adders (CLA), for faster computation. The performance of these parallel adders is significantly restricted by the longest carry chain, which is a major limitation. For an *N*-bit adder, the worst-case carry propagation is bounded by a logarithmic delay, *log(N)*, as suggested in [5]. Lu proposed an approximate adder design in [6]. This design employs the concept of *speculative* carry to minimize the carry-chain by restricting the number of bits involved in carry computation, resulting in a sub-logarithmic delay rather than a logarithmic delay. Verma et al. [7] proposed an almost correct adder (ACA), which utilizes the concept of speculative carry computation to limit the long carry-chain to optimize for speed. The design necessitates the segmentation of the original adder to multiple mulit-bit sub-adder units, where each smaller adder performs *local addition*, limiting the carry chain. In contrast to ACA, which uses partial carry propagation to increase operation speed, Mahdiani et al. [8] suggest a design called the lower-part-OR adder (LOA), which totally disregards the lower order carry bits. In LOA, an $N$-bit approximate adder is divided into two modules: an $(N-m)$-bit *accurate* Most Significant Module (MSM) that performs exact computation and an $m$-bit *approximate* Least Significant Module (LSM) that performs calculation with a reduced level of precision. The *approximate* LSM is implemented by replacing the full adder functions with bitwise-OR functions to compute the least significant bits (LSBs) of the result whereas, the *accurate* MSM is implemented with conventional exact adders, as shown in Fig. 1(a). Since the *approximate* LSM processes the less significant input bits, inaccurate computation in this unit has a lesser impact on the result than the *accurate* MSM unit, which operates on the more significant bits. An AND operation is performed on the two most significant input bits of the LSM, i.e., $A_{m-1}$ and $B_{m-1}$, to speculate the carry input signal, $C_{in}$, for the $(N-m)$-bit *accurate* MSM, as shown in Fig. 1(a). The implementation of the OR gates, in place of traditional adders, significantly simplifies the hardware and reduces energy and area utilization, while the absence of carry chain within the LSM contributes to faster computation. The improvements, however, are accompanied by a substantial loss in accuracy.

Numerous studies followed in later years that utilized and expanded upon the concept of LOA. [9] proposed the LOA without AND (LOAWA) adder, which removed the $C_{in}$ generating AND gate from LOA and fixed $C_{in}$ at '0'. [10] proposed a design that removed logic components from the *approximate* LSM, directly substituting the lower order output sum bits with one of the input signals (A or B). This evidently improves the performance; however, at the cost of a significant loss of accuracy. An error-tolerant adder (ETA) is proposed in [11], which utilizes a different approximation technique for the LSM instead of bitwise OR operations. The approximate LSM examines each bit position for the input pairs from left to right, starting from ($A_{m-1}$, $B_{m-1}$), and performs standard addition when the input bits are (0,0), (0,1), (1,0). However, when an (1,1) pair is identified, the evaluation stops and all residual output bits from the current position to the $0^{th}$ bit are set to 1. [12] further modifies ETA to improve accuracy by introducing additional logic. In [13], Dalloo et al. presents an alternative design by modifying the original LOA, known as optimized lower-part constant OR adder (OLOCA). In OLOCA, the LSM is further divided into two segments of width $(m-k)$ and $k$-bits. The OR gates are employed to process the $(m-k)$ most significant input bit pairs, similar to LOA. The remaining $k$ least significant sum bits are assigned a constant value of '1', as shown in Fig. 1(b). [14] presents an enhanced version of LOA that incorporates a hybrid error reduction scheme (HERLOA), prioritizing the accuracy of approximate computation. As shown in Fig. 1(c), additional logic elements are added in the LSM to dynamically control the approximation based on input patterns. The design substantially improves the accuracy in comparison to LOA; however, this is achieved at the expense of energy and area overhead. A modified HERLOA design (M-HERLOA) is introduced in [15], which adopts the OLOCA approach in the lower section, as shown in Fig. 1(d). This leads to a reduced hardware complexity relative to HERLOA, while maintaining comparable accuracy.

III. PROPSED APPROXIMATE ADDER

In this section, we present the design and operation of our proposed approximate adder. The design draws inspiration from the static approximate adders that were previously discussed in Section II and is further built upon to obtain a better balance between computational accuracy and efficiency for an image processing application.

*A. Design of the Proposed Adder*

The proposed design, Half-Adder Lower-part-OR with Constant Approximate Adder (HALOC-AxA), is illustrated in Fig. 2, which is obtained by modifying the *approximate* LSM logic in M-HERLOA. The goal is to further improve the design for area, speed, and/or energy utilization, while maintaining the adder's flexibility and adaptability to suit different application requirements. Like previous designs, the proposed N-bit adder consists of an *(N-m)-bit accurate* MSM and *m-bit approximate* LSM. The target application's tolerance level to computational inaccuracy must be carefully considered when determining the width of the LSM (*m*). The *accurate* MSM can be implemented using any conventional adder, such as the ripple carry adder (RCA) or CLA. The *approximate* LSM is further divided into two parts: a *k*-bit lower section and an *(m-k)*-bit upper section, as depicted in Fig. 2. The lower section employs the same methodology as M-HERLOA or OLOCA, in which the *k* least significant sum bits are directly connected to *Vdd*, forcing bits $S_{k-1}$-to-$S_0$ to be logic 1. The *(m-k)*-bit upper section is designed using a two-step approach: i) bitwise OR operations are performed on the input pairs from $(A_{m-3}, B_{m-3})$ to $(A_k, B_k)$ to produce the sum bits from $S_{m-3}$-to-$S_k$, as per LOA; and ii) two half-adders are introduced that operate on the two most significant input bit pairs of the LSM, $(A_{m-1}, B_{m-1})$ and $(A_{m-2}, B_{m-2})$. The carry generated from the addition of $(A_{m-2}, B_{m-2})$ is propagated to predict the correct $S_{m-1}$ bit, while the carry generated by adding $(A_{m-1}, B_{m-1})$ is supplied as the *Cin* input to the *accurate* MSM, as illustrated in Fig. 2.

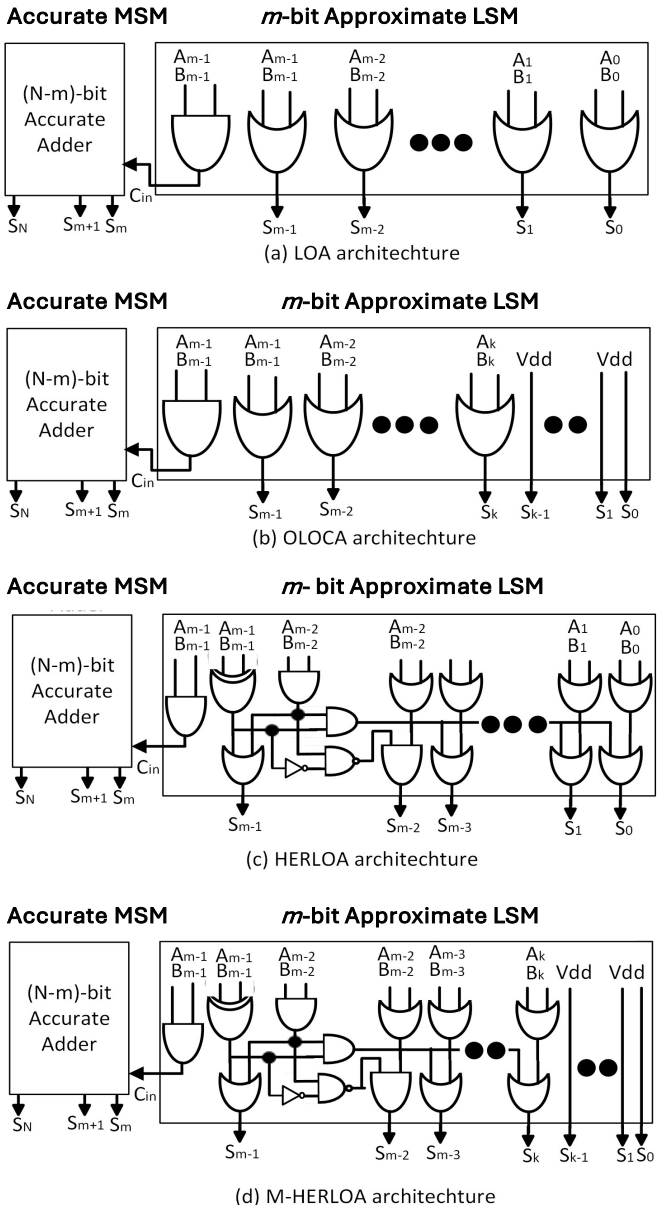

Fig 1: Existing approximate adders.

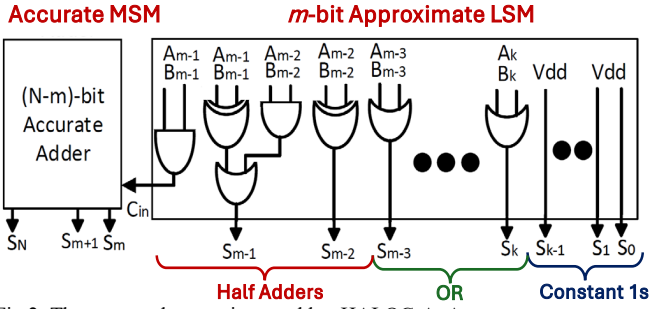

Fig 2: The proposed approximate adder, HALOC-AxA.

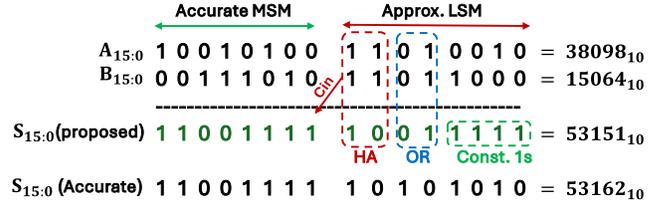

Fig 4: An example to illustrate the operation of the proposed adder.

### B. Operation of the Proposed Adder

It is straightforward to understand that the MSBs of the *approximate* LSM play a more crucial role in determining the accuracy of the overall adder. The intention behind integrating the half adders for the two MSBs of the *approximate* LSM was to effectively extend the accurate addition by two bits, without introducing substantial hardware overhead. Herein, we examine the manner in which the two MSBs are addressed in various designs and compare their influence on accuracy. Among the two bits of A, $[A_{m-1:m-2}]$, and two bits of B, $[B_{m-1:m-2}]$, there are 16 possible combinations, out of which 6 are redundant due to the commutative nature of the addition operation (i.e., *A+B = B+A*). Therefore, we consider the 10 non-redundant combinations and evaluate the summation operations of the two MSBs of the *approximate* LSM based on LOA, HERLOA, and the proposed adder, as shown in Fig. 3. Note that M-HERLOA is not depicted in Fig. 3 because the corresponding sum bits in M-HERLOA are computed in the same manner as HERLOA. It can be observed that LOA has an error rate of 50%, performing incorrect computations for 5 of the 10 combinations. In contrast, both HERLOA and HALOC-AxA perform only one incorrect computation that occurs when both $A_{m-2}$ and $B_{m-2}$ are 1 and $A_{m-1}$ and $B_{m-1}$ are different, reducing the error rate to 10%. Even though both HALOC-AxA and HERLOA computed the same combination incorrectly, the result produced by HERLOA is closer to the accurate value, indicating a minimized error distance for certain combinations. Also note that the carry propagation logic implemented between the two MSBs of the *approximate* LSM is essential in the proposed design. The absence of the logic would further increase the error rate of the proposed adder, generating erroneous outputs when $[A_{m-1:m-2}]$ and $[B_{m-1:m-2}]$ are both 01 or 11. Fig. 4 illustrates the complete operation of a 16-bit addition using HALOC-AxA. Assume that both the *accurate* MSM and *approximate* LSM comprise 8-bits each, i.e., *N=m=8*. *k=4*, indicating that the last 4-bits of the sum output, $S_{3:0}$, are all '1'. The two MSBs of the *approximate* LSM's sum output, $S_{7:6}$, are computed with the two half-adders. Finally, the remaining sum bits of the *approximate* LSM, $S_{5:4}$, are produced through the OR gates. For the same set of values for A and B, as shown in Fig. 4, the error distance (ED) of the proposed adder, which is defined as the absolute difference between the approximate and accurate outputs ($|S_{accurate} - S_{inaccurate}|$), is computed to be just $11_{10}$ ($53,162_{10}-53,151_{10}$), indicating that the output very closely approximates the accurate sum.

### IV. SIMULATION RESULTS AND DIGITAL IMAGE PROCESSING

Five existing 32-bit approximate adders (LOA, LAOWA, OLOCA, HERLOA, and M-HERLOA) and an accurate CLA were designed to evaluate their performance against the proposed design. Each approximate adder was designed with a 10-bit *approximate* LSM (i.e., *N=32, m=10*) for fair comparison. This is consistent with [15] and [16], which demonstrate that such partition achieves an acceptable balance between performance and accuracy in image processing applications. Furthermore, for OLOCA, M-HERLOA, and the proposed design, the least significant 5-bits are considered for the constant part (i.e., *k=5*), which is in alignment with [15]. All the adders were implemented at the transistor level using

| A | 00 | 01 | 01 | 10 | 10 |
|---|----|----|----|----|----|
| B | 00 | 00 | 01 | 00 | 01 |
| Accurate | 000 | 001 | 010 | 010 | 011 |
| LOA | 000 | 001 | 001 | 010 | 011 |
| HERLOA | 000 | 001 | 010 | 010 | 011 |
| HALOC-AxA | 000 | 001 | 010 | 010 | 011 |

| A | 10 | 11 | 11 | 11 | 11 |
|---|----|----|----|----|----|
| B | 10 | 00 | 01 | 10 | 11 |
| Accurate | 100 | 011 | 100 | 101 | 110 |
| LOA | 110 | 011 | 011 | 111 | 111 |
| HERLOA | 100 | 011 | 011 | 010 | 110 |
| HALOC-AxA | 100 | 011 | 010 | 010 | 110 |

Fig 3: Handling of the 2 MSBs of the approximate LSM in different adders

Synopsys HSPICE with the high-performance 32nm CMOS predictive technology model (PTM) standard cell library [17]. Several critical performance metrics, including transistor count, avg. switching power, avg. switching latency, and avg. switching energy per operation, were evaluated for each adder. A comprehensive error analysis was performed on the various adder architectures, employing 10 million ($10^7$) random input patterns. The accuracy of each adder was evaluated using two critical metrics: mean error distance (MED) and mean relative error distance (MRED), which are defined as follows:

$$MED = \frac{1}{n}\sum_{i=1}^{n} ED_i \; ; \; MRED = \frac{1}{n}\sum_{i=1}^{n} \left| \frac{ED_i}{S_{i,accurate}} \right|$$

where $n$ represents the number of input combinations, $ED_i$ denotes the error distance for the $i^{th}$ inputs, and $S_{i,\,accurate}$ refers to the accurate sum for the $i^{th}$ input combination. A MATLAB model was developed and used for this analysis.

Table 1 presents the simulation results. The findings indicate that the proposed adder necessitates fewer transistors than all other adders, except for OLOCA, which requires 24 (1.6%) fewer transistors, and LOAWA, which has the same transistor count. The switching delay of all the adders are almost identical. The proposed adder demonstrates the lowest energy utilization during switching than all other adders. In terms of accuracy, the proposed design fares significantly better than LOA, LOAWA, and OLOCA. HERLOA and M-HERLOA are both more accurate than the proposed adder; however, they require a greater number of transistors and consume more energy during switching.

To evaluate the performance of the proposed adder and compare it with that of the accurate and other approximate adders, we implemented them in an image processing application. For this purpose, a random digital image was used from [18], with an 8-bit grayscale resolution and a 512×512 spatial resolution, as illustrated in Fig 5. The image reconstruction process involves performing Fast Fourier Transform (FFT) and inverse FFT (IFFT) operations. Fig. 5(a) illustrates a digitally reconstructed image that was generated using precise multipliers and adders, performing accurate FFT and IFFT computations. Figs. 5(b)- 5(f) depict the same image reconstructed using accurate multipliers but *approximate* adders (LOA, OLOCA, HERLOA, M-HERLOA, and HALOC-AxA). The quality of the images is assessed based on two parameters: Peak Signal-to-Noise Ratio (PSNR) and the Structural Similarity Index Measure (SSIM). The structural similarity between a reconstructed image and an original image is measured by SSIM, while PSNR serves as a general benchmark for digital signal processing quality. SSIM > 0.90 is considered high-quality, $0.70 < SSIM \leq 0.90$ is deemed acceptable, $0.30 < SSIM \leq 0.70$ is regarded as low-quality, and SSIM < 0.30 is categorized as poor quality. Fig. 5 indicates that the proposed adder yields high-quality images (SSIM= 0.92), while LOA and OLOCA produce images of acceptable quality (SSIM=0.85). Note, LOAWA further degrades the image quality (SSIM=0.75); however, the reconstructed image is omitted owing to the page limitations. Among all the adders, HERLOA and M-HERLOA produced images with the highest quality (SSIM= 0.94), which is consistent with our previous analysis. Fig. 6 depicts the trend in SSIM values and normalized switching energy consumption per operation for all the adders. The plot indicates that the proposed adder consumes the lowest average switching energy per operation, while producing high-quality images that are comparable to HERLOA and M-HERLOA.

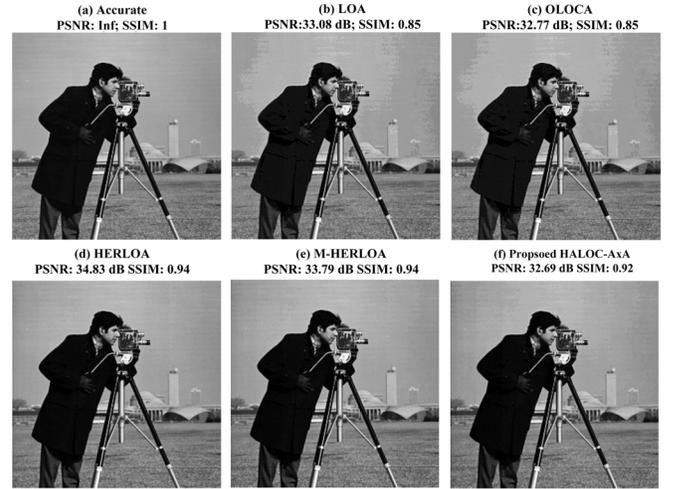

Fig 5: Image reconstruction employing (a) accurate adder, (b) LOA, (c) OLOCA, (d) HERLOA, (e) M-HERLOA, and (f) the proposed adder.

Table I: Simulation Results

| Adders | Trans. Count | Avg. Switching Power (μW) | Switching Delay (ns) | Avg. Switching Energy (fJ) | MED | MRED (×10⁻⁸) |
|---|---|---|---|---|---|---|
| Accurate | 2208 (+43%) | 302.19 (+33.5%) | 0.24 (+9.5%) | 66.25 (+29%) | N/A N/A | N/A N/A |
| LOA | 1548 (+0.4%) | 242.18 (+7.0%) | 0.21 (0%) | 55.05 (+7%) | 191.9 (+55%) | 6.19 |
| LOAWA | 1542 (0%) | 237.86 (+5.1%) | 0.21 (0%) | 53.42 (+4%) | 255.7 (+106%) | 8.25 |
| OLOCA | 1518 (-1.6%) | 226.69 (+0.1%) | 0.21 (0%) | 51.71 (+0.5%) | 190.6 (+54%) | 6.15 |
| HERLOA | 1632 (+5.8%) | 265.15 (+17.1%) | 0.21 (0%) | 60.04 (+17%) | 97.7 (-21%) | 2.94 |
| M-HERLOA | 1572 (+1.9%) | 233.57 (+3.2%) | 0.21 (0%) | 52.92 (3%) | 94.9 (-23%) | 2.91 |
| **Proposed** | **1542** | **226.39** | **0.21** | **51.45** | **123.9** | **3.77** |

(+)/ (-) indicates relative increase/ decrease compared to the proposed adder

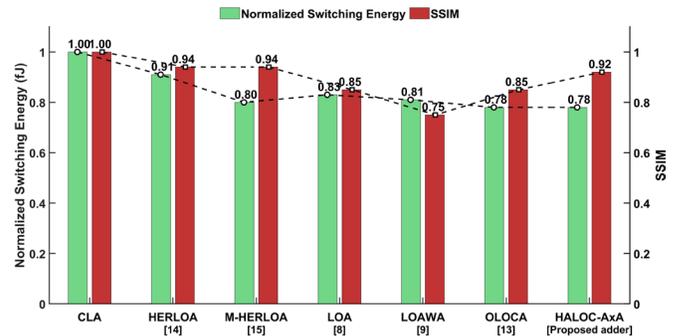

Fig 6: The trend in SSIM and normalized switching energy.

## V. Conclusion

In this paper, we present a novel approximate adder, HALOC-AxA, suitable for multimedia applications, such as image/audio/video processing. The simulation results indicate that the proposed design is more energy- and area-efficient than existing adders. The design's ability to reconstruct high-quality images was further demonstrated by the deployment of the adder for an image processing task, which achieved superior or comparable accuracy in comparison to existing approximate adders.